\documentclass[aps,prd,superscriptaddress,eqsecnum,amsfonts,showpacs]{revtex4}
\usepackage{graphicx}

\newcommand{\be}{\begin{equation}}
\newcommand{\ee}{\end{equation}}
\newcommand{\bea}{\begin{eqnarray}}
\newcommand{\eea}{\end{eqnarray}}

\newcommand{\nn}{\nonumber}
\newcommand{\ep}{i\epsilon}
\newcommand{\om}{\omega}

\begin{document}

\title{A study of spin-one bottomonium   in the functional formalism in Feynman gauge}

\author{V. \v{S}auli  }
\affiliation{Department of Theoretical Physics, Institute of Nuclear Physics Rez near Prague, CAS, Czech Republic  }

\begin{abstract}
The spin-one bottomonium  Bethe-Salpeter equation (BSE) is solved in a class of nontrivial linear gauges. 
The self-consistent ladder-rainbow approximation, based  on the effective coupling extracted   from the lattice,
 lies at the core of   the considered approximation.  The  first numerical results  obtained in the Feynman gauge
 show that a very little phenomenology is needed. We address the issue of infrared divergences in the functional 
formalism for a nontrivial gauge-fixing parameter.

\end{abstract}

\maketitle

\section{Introduction}

 This paper focuses on calculating the spin-one bottomonium spectrum using the fully covariant Bethe-Salpeter equation.  
We use a ladder-rainbow kernel composed of known gluon Green's functions. It is well-established that the Landau gauge improves ultraviolet behavior while avoiding infrared divergences due to the mass generation effect in a completely transverse gluon propagator.   
However, the Landau gauge is not considered in isolation, despite its reputation in practice. We work in the class of nontrivial covariant gauges and address the broader issue of infrared divergences arising from the gauge term in the interaction kernel.
Despite the issues associated with the massless limit of the gauge-fixing term, our preliminary investigation indicates that, within the Feynman gauge, a minimal amount of phenomenology is necessary to adequately describe quarkonia below the $BB$ threshold. 

In light of the limited number studies of the quarkonium BSE, a review of alternative methods for addressing 
the bottomonium system is warranted. 
For topical quarkonia rewiev see \cite{BRAPRO1990,BAPR1998,EGMR2008,EFG2011,B2011,PPR2012,CH2022}.

\subsection{Quantum Mechanical $Q\bar{Q}$ models}

 Excited quarkonia serve as an exemplary laboratory for exploring a myriad of theoretical concepts.
In the  early stages of the history, Regge trajectories were found to exhibit only negligible deviations 
from linearity $M^2(n)\simeq n$. 
Quantum mechanics has provided numerous microscopic-level explanations.The Lorentz vector and/or scalar,  
enough  severe  infrared  interaction of the type $\simeq 1/q^4$,  has been widely accepted as a prototype   
for the confinement of quarks \cite{KS1974,KS1975,QR1977,EGKLY1980}. At the time of its initial formulation, the associated string theory postulated
 the inverse quartic interaction rather than the  quadratic    propagator form associated with a possible   exchange of   vector bosons.
 Subsequently, when QCD was taken more seriously, a one gluon $1/q^2$ interaction was incorporated into the  semiperturbative interaction between  quarks. 
While it has been incorporated into the potential, it has not yet replaced the confining part. 
The non-relativistic $X-space$ version  that consists from  Coulomb term
and a kind of confining interaction is used in plethora of its original extensions.
 Actually, at these days we are faced a plethora of 3-d potentials suited for particular phenomena descriptions, light-front quark models 
\cite{LMZZV2016,BPSS2024,RAM2025} was found to be particularly universal.  Further potential models \cite{MSRS2020,KCDR2022,SLM2024} and more evolved 
effective theory coupled channel analysis \cite{LD2012,BCMW2022}  found their applications bellow and above $BB$ production threshold respectively.
 
Here we assume, that in the four dimensional formalism of Dyson-Schwinger and Bethe-Salpeter equations the confinement   
is inherently embedded within the infrared behaviors of running couplings.  It is possible to formulate these in a 
nonperturbative manner by considering combinations of Green's functions that are renormalization invariant. 
The selection of an appropriate computational scheme, inclusive of the choice of gauge fixing, if applicable, 
has the potential to enhance or diminish the visibility of this representation. 
The aforementioned factors serve to further motivate the present study.

\subsection{QCD functional formalism of gap equations}   
The Dyson-Schwinger equations (DSEs)  \cite{DSE1,DSE2,DSE3}  are equations of motion for Green's functions.
They are used to formulate continuous form factors in the strong coupling region and are incorporated into bound state calculations. 
In the standard approach to the theory, gauge fixing must be implemented as a preliminary step to use the formalism. 
Although observables are expected to be gauge-invariant, the solvability of the system and the quality of solutions are matters of concern.  
Selecting the most suitable gauge for a given problem reduces the complexity of the treatment.

Note that the known quarkonium BSE studies
 were  employed in the absence of purely longitudinal interactions; i.e.   this term was either ignored or   they were  calculated in  Landau gauge.
 In \cite{BLKR2011, VS2012, VS2014, GKK2018}, a phenomenological inputs, such as a regularized linear confining potential or a Maris-Tandy Gaussian interaction,
 were added to the ladder-rainbow kernels.  
To achieve similar goals in Landau gauge, the concept of the effective running quark mass  was employed in \cite{vs2025} , 
has also been shown to work. To further address this unexplored issue further, we restrict ourselves to the 
ladder-rainbow approximation (LRA) and check nontrivial gauges  to minimize the strength of the phenomenological input. 
After some testing, we found that the infrared-regularized Feynman gauge is particularly suited for this purpose.
 For practical reasons, we will present the results in this gauge.

For other mesons as well, the LR, which is limited to 
$\gamma^{\mu}\times\gamma^{\mu}$  matrix structure, always requires an effective interaction
\cite{MARO1997,MATA1999,MATA2000,JAMATA2003,MATA2002,MATA2006,KRA2009,KRBL2011,CCRS2013}. To eliminate phenomenological kernels, 
one must improve the truncation method and extend beyond the LR, as discussed in 
\cite{BCC2012,WCT2004,BHK2004,FWC2005,MTT2007,AFW2008,FW2009,CR2009,HGF2014,VW2014,W2015,WFH2016,CMP2018} 
   in order to include  the correct strength of QCD vertices.
Studies in Landau gauge dominate other QCD sectors as well \cite{Bar2021,P2019,SFAP2020,EF2018}. 
The Pinch Technique  \cite{CPB2011,FP2023},  which 
advocates  Feynman Background gauge    in  DSEs calculations, is still waiting for its first application in  hadron physics.

The Schwinger mechanism  (see, for instance, \cite{ABP2017}) 
has been proposed as a way to generate gluon mass in a gauge-invariant manner.  It causes severe infrared divergences to be quenched. Consequently, 
infrared divergences are essentially absent in the quark-gluon sector when the Landau gauge is used. 
However, infrared divergences reappear in other gauges, creating difficulties with composite vertices that describe hadrons. 
This includes quarkonia as well.
Clearly, these divergences are not canceled by real soft-emitted gauge bosons because the Kinoshita-Lee-Nauberger theorem does not apply here.
Thus, nonperturbative studies in nontrivial covariant gauges open a Pandora's box 
of nonperturbative applications of the Faddeev-Popov quantization method \cite{FP1967,gribov1978}.

\section{DSE and BSE  for spin one quarkonia}

In this section, we review the set  equations actually used to evaluate quarkonia properties. 
Note that the preliminary versions of the presented DSEs truncations were already used to describe  the   properties of pions 
\cite{VS2020} and pseudoscalar charmonium \cite{VS2024}.

Let us start with the DSE for the quark propagator, which 
reads
\be
 S^{-1}(p)=Z_2\not p -Z_m m -\Sigma(p) \, ,
\ee
\be
\Sigma(p)=\frac{4}{3}\int_q  \gamma S(q)  \Gamma_{gq}(q,p)  G(q-p) \, ,
\ee 
where the integration symbol $\int_q$ in expression for the quark self-energy $\Sigma$ stands for the 4dim integration measure 
$i\int\frac{d^4q}{(2\pi)}^4$ and we have suppressed all Lorentz, Dirac and color indices for brevity.

The interaction of the quark-antiquark pair inside the meson and the running quark mass could  
originate from the a meaningful symmetry preserving truncation of the DSEs system.
To mainatin selfonsistence we will work in the  Ladder-Rainbow approximation (LRA), 
which means that the interaction kernels used in the quark DSE is identical to the kernel
of the BSE for bound state. The selfconstince becomes more  aparent in case of vanishing quark mass
where a given approximation ensure the equivalence of pseudoscalar bound state (the pion) with Goldsteone boson.

To state the  more explicitly, the LRA  consists  from the  quark-gluon vertex  $\Gamma_{gq}^{\mu}(q,p)$ 
solely  proportional to  $\gamma^{\mu}$ dirac matrix and the gluon propagator, which reads
\be \label{gluon}
G^{\mu\nu}(k)=(-g^{\mu\nu}+\frac{k^{\mu}k^{\nu}}{k^2})G(k^2,\xi)-\xi\frac{k^{\mu}k^{\nu}}{[k^2]^2}\, .
\ee

The renormalization of the quark DSE is  standard and we use dimensional regularization to regularize it,
which  immediately leads to DSE in dimensional renormalization scheme.
To obtain the solution in the entire Minkowski space, we renormalize again using a finite subtraction. This yields   the momentum-space renormalization scheme with the timelike renormalization scale $\mu$. The subtracted  terms are conveniently  absorbed into the counterterm parts of the  $Z_m$ and $Z_2$ renormalization constants. The renormalized DSE then reads

\be
 S^{-1}(p)=Z (\mu)\not p - m(\mu) -\Sigma(p,\mu) \, ,
\ee
where the renormalized function $\Sigma $ can be decomposed into two scalar functions $a$ and $b$:
\bea
\Sigma(p,\mu)&=&\not p a(p,\mu)) +b(p,\mu))\, ;
\nn \\ 
a(p,\mu)&=&a(p)-a(\mu) \, ;
\nn \\
 b(p,\mu)&=&b(p)-b(\mu)\, ,
\eea
where for notational convenience, we have introduced single-scale dependent functions $a(x)$ and $b(x)$.
The parts that stem solely from the gauge term (the second term in the Eq. (\ref{gluon}) have particularly simple forms. They read
\bea \label{new}
b_{\xi}(p) &=& \lambda \int_0^{p^2} dq^2 (1-r) S_B(q^2) \, ;
\nn \\
a_{\xi}(p)&=& \lambda \int_0^{p^2} dq^2 (-r^3+2r^2-1) S_A(q^2) .
\eea
where we denote $r=q^2/p^2$ . Further, the coupling reads 
\be  \label{coupl}
\lambda= C_A \frac{g^2}{(4\pi)^2}
\ee
with Casimir value $C_A=4/3$ for $SU(3)$. Two introduced scalar  functions $S_A$ and $S_B$ 
complete the quark propagator,  defined as:
\be \label{kumbucha}
S(p)=\not p S_A(p^2)+ S_B(p^2)\, .
\ee 

The derivation of Eq. (\ref{new})  is  delegated into the   appendix.
To derive the remaining part of the self-energy, we used well-known dispersion relations. 
We extracted the two quark spectral functions,  which are defined as follows:
\be  \label{SF}
\sigma_A(p)=-\frac{\Im S_A(p^2+\ep)}{\pi}\,\, ;\,\, \sigma_B(p)=-\frac{\Im S_B(p^2+\ep)}{\pi} \, .
\ee
from the solution of the quark DSE. 
This process is detailed in references \cite{VS2020, VS2022, VS2024}, which employ similar methods to those used in references \cite{HOPAPAWI2021, HOPAWI2023} for the same purpose. The obtained spectral functions $\sigma_{A,B}$ provide a smooth analytical continuation of the quark propagator into the complex plane at the necessary points for evaluating the BSE.  
 
In the case of the DSE solution, the nonproblematic kinematical infrared singularity can occur if the spectral function is nonvanishing at the origin.
However, in the BSE solution, the kernel singularities are more problematic.
To ensure consistency with LR, we solve the BSE with the same kernel as for the quark DSE. The BSE sketch for spin-one quarkonia is as follows: 
\be
\Gamma_{BSE}^{\alpha}(p,Q)= \frac{4}{3}g^2 \int_q \gamma^{\mu}  S(k_+)\Gamma_{BSE}^{\alpha} (k,Q) S(k_-)
\Gamma_{gq}^{\nu}(k,q)G_{\mu\nu}(q-p) \, .
\ee

Here, the symbol  $\Gamma_{BSE}^{\alpha}$ stands for the Bethe-Salpeter quarkonium vertex function, It is a $4\times4$ matrix with dirac indices (not shown). The Lorentz index $\alpha$ is due to the spin one.  Of the eight transverse components, we begin with the first and fifth components. For vectors they are usualy defined us 
\bea   \label{frnat}
\Gamma_1^{\alpha}(q,Q)&=&(\gamma^{\alpha}-\frac{\not Q Q^{\alpha}}{Q^2})\Gamma_1(Q,q)
\nn \\
\Gamma_5^{\alpha}(q,Q)&=&(q^{\alpha}-\frac{q.Q \,Q^{\alpha}}{Q^2})\Gamma_5(Q,q)
\eea 
where $\Gamma_i(Q,q)$ that appears in the rhs (\ref{frnat}) are Lorentz scalars. 
The first component is known to be dominant in related calculations in Landau gauge, the second component could vanish for $s-$wave quarkonia. The latter is used primarily to check consistency, as it does not substantially improve the studied case. The BS components for axial vectors can be defined analougously, 
differing by  presence of  $\gamma_5$ matrix in  the expression.

After the projection, the coupled equations for the aforementioned components are obtained and solved in the Euclidean space in the meson's rest system.
Although the arguments are complex, such that $p.Q=ip_4 M; -p^2=p_E^2>0,  Q^2=M^2>0$ ,$M$ is a meson mass, the  two  BS components used can be chosen purely real for vector quarkonia.  The first component can be taken  real for axial vectors, while the fifth component is then purely imaginary.

We solved the BSE numerically, simultaneously solving the gap equation for the b-quark. To reduce the number of integrations, we integrate over the spacelike angle analytically. To prevent infrared singularities in the individual terms, we introduced an infrared regulator, $m_{\epsilon}$, to the gauge term.
\be
\xi\frac{k^{\mu}k^{\nu}}{[k^2]^2}\rightarrow \xi\frac{k^{\mu}k^{\nu}}{[k^2-m_{\epsilon}^2]^2}\, .
\ee

The rest of the interaction kernel is fully specified by the transverse part of the gluon propagator.
For this purpose, we use  a simple extrapolation of the known Landau gauge  gluon form factor $G$
as extracted from the lattice \cite{HPRTU2021}, and DSEs analyses \cite{VS2022}. To reduce the number of numerical integrations, we use a two-poles approximation for the gluon form factor.
\be \label{twopole}
G(q^2,\xi)=\frac{r_1}{q^2-m_{g1}^2}-\frac{r_2}{q^2-m_{g2}^2}
\ee

It reduces the  two broad continuous peak structures observed in studies \cite{HPRTU2021},\cite{VS2022} 
into a delta functions. We take the residue and pole positions to be $r_1=1.5$ and $r_2=1$ and
 $m_{g1}^2=0.14 GeV^2$ and $m_{g2}=1.0 GeV$. Note that, as was recently observed in \cite{vs2025}, the mass scales $m_{g1,g2}$  change dramatically, when the running of the
QCD charge is incorporated into the kernel. Unfortunately, the related improvement of the BSE kernel does not achieve the necessary the  numerical precision when we stay limited by the method used in the presented paper. We do not use it here. 

\section{Taming of Infrared Divergences in Quarkonium BSE Solutions}

 The BSE components are projected onto coupled equations for individual components and solved numerically. 
It is advantageous to normalize the solution within the auxiliary function N, which is used as an eigenvalue in front of the BSE. In the case of constituent mass approximation, the search reduces to an eigenvalue problem, while the method is extended to LR approximation, where the BSE requires knowledge of the quark propagator. To this end, the quark DSE was solved in advance in a separate iteration subcycle.  
 The quark propagator is renormalized such that $Re B(5 GeV)= 5 GeV$.   In our MOM scheme, we set the value $Re A(5GeV)=1$, noting that the functions $A$ and $B$ are complex for timelike momentum and they receive imaginary parts  $\Im A(5 GeV)=2 * 10^{-5}$ $\Im B (5 GeV) =-0.724 GeV$ at the timelike renormalization scale $\zeta= 25 GeV^2$.
The spectral deviation as defined in \cite{VS2022,VS2020} takes the value   $\sigma2\simeq 3.4 * 10^{-4}$ for the  DSE solution. The bottom quark spectral functions are shown in the Fig. \ref{zuva}.

 We use iterations to solve the BSE. 
Although time-consuming, the iteration method works reasonably well with nontrivial gauges, \cite{VS2020,VS2024}.
These studies were devoted to the spin zero bound state,   where the trace projection of the BSE  reduces the gauge term singularity $1/q^4$  to $1/q^2$ for the dominant BSE component.   
 For now, however,  the structure of the $J^{PC}=1^{--}$  BSE contains unremovable infrared divergences. Note that  the BSE has an identical structure to   
the DSE  for  the quark-photon proper vertex in QED (dressed by QCD loops). The gauge proper vertex contains log infrared divergences irrespective of the quark propagator structure. Neither quark nor gluon mass can completely prevent infrared singularities because they also arise from the gauge term (of course, IDs are more severe in the presence of massless particles, but we consider this case to be unphysical since confinement and chiral symmetry breaking occur in QCD). There is little doubt that IDs can be cured by the soft properties of the bound state vertex function. They would all have to be zero at the origin of loop momentum.  Without a canceling mechanism, the BSE for vectors has no apparent physical solution for a nontrivial gauge-fixing parameter.  In our case, this could be seen as evidence of the absence of a vanishing mass regulator limit.  

{\mbox{}}
\\
\begin{figure}[ht]
\centerline{{\includegraphics[width=7.50cm]{verynice.eps}}{\mbox{-}\mbox{-}}
{\includegraphics[width=7.50cm]{verynice2.eps}}}
\caption{\label{zuva} Left panel: Bottom quark spectral function from DSE and its fit as described in the appendix. Right:
The same as in the previous figure but in log scale for $p^2$.}
\end{figure}
{\mbox{}}
\\
Note that there are specific gauges in which radiative corrections are free of IDs. The Yennie gauge ($\xi=3$), as suggested  in  \cite{EISH2001}, could be particularly useful for QED bound state studies. However, no nontrivial gauge choice renders the BSE infrared safe. 
 Our study is perhaps the first, albeit primitive, numerical attempt to understand infrared safety in the functional framework of DSE/BSE equations.  
How are the IDs seen in practice in BSE evaluations?  If there were no singularities, the infrared regulator could be kept smaller than any other scale, and one could achieve results of a similar quality to those in the popular Landau gauge or those in ID-free BSEs for pseudoscalars \cite{VS2020,VS2024}.  
 This is not what we observe. We will mention a few peculiar inconsistencies that arise as the regulator $m_{\epsilon}$ approaches its small limit. Once the numerical value of $m_{\epsilon}$ falls below a certain threshold, the interaction strength increases, resulting in a continuous solution for the bound state mass. 
 This is striking numerical effect since bound states are formed for any value of the total momentum. It is an overbinding effect. This is not obviously  observed in quarkonium production experiments; rather, we observe several very narrow, well-separated bound states instead.
 This pathology was first discovered in \cite{GOLD1953} and later confirmed 
  in more general case in  \cite{AA1998} for the systems with excessive strong coupling.
We assume this  could be prevented in  QCD, especially for bottomonia.
Another discrepancy appears in the DSE + BSE coupled scenario, even for moderate coupling values. It is impossible to obtain a simultaneous solution to the BSE and DSE equations for small $m_{\epsilon}$ . This behavior is not directly related to the numerical stability of the BSE or DSE systems, but rather to the unphysical picture that the equations reveal under poorly defined circumstances.
 
For the sake of completeness, we should note that  the Kinoshita-Lee-Nauenberg theorem \cite{KLN1,KLN2} does not prevent  from described infrared catastrophe.
In the case of a bound state, since there is no room for state  degeneracy by adding on shell radiation of colored soft gluon into the system. Confinement does not allow this. To make covariant gauges applicable, at least to some extent, we  allow the scale $m_{\epsilon}$ to be of order 
$\Lambda_{QCD}$ and search for the BSE solution.  Only then can we identify the ground and first excited $Y$ mesons.

Non-monotonous behavior  appears in the QCD vertices  
 beyond LR employed here \cite{W2015} and we  we found useful to implement  a small - about $ 15\% $ of the total strength-  phenomenological interaction and allow  oscillation 
of the coupling  in the following form:
\bea \label{osci}
&&\alpha_{S}\rightarrow \alpha_{S}(1+\delta)
\nn \\
\delta&=&\delta_0 \sin\left(2\pi \frac{Q-\phi_M}{1.12 }\right) 
\eea
 where $Q$ is the magnitude  of total momentum and $\phi_M$ is the constant phase in units of $GeV$  and the inequality $\delta_0<<1$ should be kept. We take the value $\phi_M=M_{Y(1)}=9.460$ in this paper, i.e. as there would be no fluctuation at total energy equal to the lowest well established spin one state. We found that  a small oscillating interaction makes the BSE solution particularly stable.
Notably, there is an experimental evidence  of the suggested oscillating  behavior seen for the timelike argument and described in \cite{BEDURU1980,BEDURU1983,BERU1009}.  

The vector meson spectra calculated within the LRA, as shown in Table 1, are mesmerizing. There, we compare the LRA results with Constituent BSE (CBSE) and with experimental results.
In the CBSE case, the running of the quark mass is ignored, and the quark spectral function is approximated by a single delta function.
Self-consistency of the LRA is abandoned, and the resulting constituent quark propagator is used in the original BSE. 
To check the effect of the perturbation, the CBSE was calculated with two different oscillation strengths. The first labeled as CBSE1 in Table corresponds to  $ \& \delta_0=0.15$  and the second, labeled  CBSE2, corresponds to  
 $ \delta_0=0.3$.  The Ladder-Rainbow was calculated with $\delta $ identical to the second CBSE2 case.

The coupling value (\ref{coupl}) was set to  $\lambda=5.193\, 10^{-4}$ for the CBSE1 and its value was adjusted to $\lambda=5.972 \, 10^{-4}$ for the  purpose  CBSE2 evaluation. In other words, we allow twice the fluctuation around a slightly larger mean coupling strength. To conclude our numerical observations, we should mention that lower values of the coupling ($\lambda$) result in the loss of states that we would prefer to keep. Conversely, higher values of the coupling lead to the aforementioned inconsistencies in the DSE and BSE solutions.  
As can be seen in  the  Table \ref{tabrho}, there is no good evidence for the 10360 meson (it is $3s$ state in non-relativistic QM), which may be due to the inefficiencies of the iteration/integration routine and/or the absence of the QCD running coupling in the transverse part of the interaction kernel.  
 We will leave this problem unanswered for now, expecting more complete agreement in the future, as the kindred for charmonia already suggests \cite{vs2025}.




The first component is a slowly varying function for both $q.Q=q_4M$ and $q_s$ variables. We illustrate its behavior for a fixed $q_s=0.5$ in the Fig. \ref{vobr1}. The normalization is unphysical so the eigenvalue function is used instead. The physically equivalent solution with an opposite sign is shown for comparison of the effect of numerics. The fifth component of the BSE vertex function vanishes for vectors, being as small as the numerical fluctuation of the first component. 

For the purpose of CBSE, the constituent quark mass was taken as $m_b=4$ GeV. Providing the same spectra in different approximations, the constituent quark mass was found to be significantly smaller than the maximum peak position in the bottom quark spectral function.
\\
{\mbox{-}}
\\
\begin{figure}[ht]
\centerline{{\includegraphics[width=7.0cm]{wave0.eps}}{\mbox{-}\mbox{-}}
{\includegraphics[width=7.0cm]{wave.eps}}}
\caption{\label{vobr1} $\Gamma_1(q_4,q_s=0.5GeV)$ for several vectors  (left) and $\Gamma_5(q_4,q_s=0.5GeV)$ component for the axialvectors (right).
 Auxiliary normalization condition \ref{normalization} is used.}
\end{figure}
\\
{\mbox{-}}
The regulator was fitted in the  LR  and simply reused in the CBSE approximation. Its numerical value is therefore
\be  \label{master}
m_{\epsilon}^{LR}= m_{\epsilon}^{CBSE}= 374 MeV  \, .
\ee
Its value prevents inconsistencies in the BSE solution. Since it is as large as the gluonic kernel parameter 
$m_{g1}$,  it becomes effectively becomes  part of the interaction kernel.

 \begin{center} 
\begin{table}\label{tabulka}
\begin{tabular}{ |c|c|c|c|c|}
\hline
\hline
 name & LR DSE     &  CBSE1  &CBSE2 & Experiment   \\
\hline
$Y(1)  $ &  9.48     & 9.510  &  9.437      &  9.460       \\
$Y(2)$   & 10.00      & 10.00  & 10.075     & 10.026      \\
$Y(3)$   &   ?        &  ?   & ?          &  10.3778     \\
\hline
 $Y(4)$  &10.61     & 10.64  &10.559      & 10.579  \\
 $Y(5)$  & 11.15    & 11.07 &  11.155     &  10.753     \\
 $Y(6)$  & 11.72    & 11.45* &  11.698     &   10.860    \\
 $ Y(7)$ & 12.26    & 11.77 &  12.610      &   11.020    \\  
 \hline
 \hline
\end{tabular}
\caption{\label{tabrho}Comparison of Constituent BSE,  the  dressed LR  and  PDG data as described in the text.
  The horizontal line marks open bottom $BB$ thresholds.  We included the Belle observed state 10753.
 States have spin one and parities $P,C=-1 $.   }
\end{table}
\end{center}

 
 \subsection{Axial vector bottomonia}
 
  We searched for the $h_b(n)$ spectra by reflecting the changes in  propagator products in  the BSE  according to the bound state quantum numbers.  The BSE interaction kernel was taken to be exactly identical to the vector BSE in the considered approximations. 
 Since the interaction structure differs beyond the LR, it is reasonable to permit a change in the constant phase in our perturbation, given by
 $\delta=\delta_0 \sin 2\pi \frac{Q-9.6}{1.12}$ where $Q$ is the meson bound state mass again. We assume the origin of the period is in the Yang-Mills part of the QCD interaction and take it to be the same as in the previous case.

We compare the results obtained for CBSE with the experimental data in the Tab \ref{bulka2}.  When the true minima of the error are considered, there are no missing states below the threshold, and the error is small $\sigma^2\simeq 10^{-5}$.   
We  show the first and the fifth  BSE   component against the relative (imaginary) energy in the right panel of the Fig \ref{vobr1}.
Unlike vectors, the  function$\Gamma_5$  is not negligible, it can  dominate over $\Gamma_1$.

  \begin{center} 
\begin{table}
\begin{tabular}{ |c|c|c|c|}
\hline
\hline
 name &     CBSE 1 & CBSE 2 &  Experiment   \\
\hline
$h_b(1)  $ &   $9783  \pm 15$ &  9970      &  9899       \\
$h_b(2)$   &  $10231 \pm 20$  & 10300     & 10259       \\
\hline
$10610$   &  $10500  \pm 20$       &  10640 & 10610       \\
 $10650$  &    $10800 \pm 20 $    &  10990     &     10650       \\
 \hline
 \hline
\end{tabular}
\caption{\label{bulka2}Comparison of CBSE and PDG averages for   $ J^{CP} =1^{+-} $ states. CBSE1 and CBSE2 solution  was obtained with $\delta_0=0.15$ and $\delta_0=0.3$ respectively as described in the text.
We expect the content of experimental states bellow horizontal line is more complex then quarkonium state and the comparison should be taken with a  care.}
\end{table}
\end{center}

 We demonstrate the sensitivity to the suggested fluctuating term and solve  the axial vector BSE with 
$\delta_0=0.15$ $\delta_0=0.3$ now for the identical coupling $\lambda=5.972$ for simplicity.  The calculated spectra are shown in the in the first and the second column of the Tab \ref{bulka2} respectively. Within a small scanning step, we observe the doubling of solutions; that is, there are two very close solutions instead of one. 
 Although the difference is comparable to the width of each state, this phenomenon is likely related to the weakness of our iteration/integration numerical method or the inadequacy of our basis choice, rather than to the physics of resonances. This effect requires further study, and we use it to estimate the systematic error in our case. We do not show the LR result for axial vectors since it plainly agrees with the CBSE result within the estimated error.

 \section{Conclusion}

In this paper, we examine the DSE/BSE system for vector and axial-vector quarkonia in the Feynman gauge, seeking a numerical solution.
 Below the $B\bar{B}$ production threshold, we observe reasonable agreement with experimental data. However, the picture becomes inadequate for describing the complex resonant structure above the threshold. The advantage of the presented calculation is that it limits the use of strong phenomenological interactions. On the other hand, reproducing the experimentally observed spectrum required using the infrared regulator mass in the gauge term. In the absence of a theory for canceling infrared divergences in the bound state vertex, it is plausible that this excludes the class of linear renormalizable gauges from nonperturbative, gauge-invariant considerations. Since we only have numerical evidence, the consistent use of linear gauges in nonperturbative calculations could be further studied.  

 This finding aligns with the anticipated observation that the dressing of the bottom quark propagator plays a non-essential role
 in determining the mass spectra of heavy quarkonia. 
Conversely, the observation of a pole-less solution for the dressed b-quark propagator indicates confinement in the heavy quark sector.
The bottom quark propagator displays a broad peak with a width of 1-GeV, which corresponds to the brief duration that the bottom quark spends in a specific color state.The washout, and consequently the absence of a quark production threshold in the hadronic form factor, is a predicted consequence.
In this regard, a high degree of similarity is exhibited between the confinement reflection in the bottom quark propagator and that observed for the lighter degrees of freedom in QCD \cite{VS2020,VS2022,VS2024}.  .

Due to numerical challenges, we were unable to employ the running coupling that incorporates the accurate UV tail as known from Perturbation Theory.
It is important to note that undertaking this action may result in certain alterations, as indicated by the initial variational solutions of a related model \cite{vs2025}.
Notably , the gluonic scales  $m_{g1},m_{g2}$  employed in the interaction kernel of our BSE gets larger to get rid of overbinding.
Actually, applying improved model \cite{vs2025} from charmonia  to bottomonia, one reproduces the ground state by 
taking  $m_{g1}\simeq 1.2 GeV$ and  $m_{g2}\simeq 5 GeV$. Further  studies of additional hadron states and their formfactors are scheduled to be conducted within the formalism that has been presented.

\appendix

\section{DSE Selfenergy evaluation}

Let us  split  the selfenergy function $\Sigma=\Sigma_t+\Sigma_{\xi}$
\bea \label{sigmy}
\Sigma_t(p)&=&i C_A g^2 \int\frac{d^4k}{(2\pi)^4} \gamma^{\mu} S(k)\gamma^{\nu}
\int d o \sigma_g(o) \frac{-g_{\mu\nu}+\frac{(k-p)_{\mu}(k-p)_{\nu}}{(k-p)^2}}{(k-p)^2-o+\ep}
\nn \\
\Sigma_{\xi}(p)&=&-i\xi C_A g^2 \int\frac{d^4k}{(2\pi)^4} 
\gamma^{\mu} S(k)\gamma^{\nu} \frac{(k-p)_{\mu}(k-p)_{\nu}}{((k-p)^2)^2}\, .
\eea 
according to the tensor structure of the gluon propagator.  We have  used the spectral representation
 to express this propagator in the Eq. (\ref{sigmy}). The function $\Sigma_t$ was evaluated  and expressed via standard
 dispersion. The results were used to solve the DSE.
 For the function $\Sigma_{\xi}=\not p a_{\xi} (p^2)+ b(p^2)$ we derive  the result in more compact form in this appendix.

Using also the following spectral representation  for the quark propagator functions (and we do not write boundaries when they are  $0$ and $\infty$ for spectral integral or over the entire $R$ for momentum integral) :
\be \label{srq}
S_{A,B}(p)=\int d \om \frac{\sigma_{A,B}(\om)}{k^2-\om+\ep}
\ee
then after the renormalization,
one gets for two scalar functions $b,a$ in ${\bar{MS}}$ renormalization scheme. 
Here we show  steps in details
\bea
b^{MS}_{\xi}(p)&=&-i\xi C_A g^2 \int_{DR}\frac{d^4k}{(2\pi)^4} 
\int d \om \frac{\sigma_B(\om)}{k^2-\om+\ep}\frac{1}{(k-p)^2}
\nn \\
&=&\xi \lambda \int^1_0 dx 
\int d \om \sigma_B(\om) ln \frac {p^2 x(1-x)-\om x +\ep}{-\mu_t^2}
\nn \\
&=&\xi\lambda \int^1_0 dx 
\int d \om \sigma_B(\om)\left[-1+ln \frac{p^2 (1-x)-\om  +\ep}{-\mu_t^2}\right]
\eea
for the mass function $b$.

At this point we will explore spectral representation (\ref{srq})  once again and rewrite  the log
in the following way:
\be
\int d \om \sigma_B(\om) ln \frac{p^2 (1-x)-\om  +\ep}{-\mu_t^2}=
\int_0^{p^2(1-x)} S_B(q^2)+\int d(\om) \sigma_B (\om) ln\frac{\om}{\mu_t^2} \, .
\ee

Interchanging the order of the integrations, one can integrate over the variable $x$, which leads 
to the following  formula:
\be
b^{MS}_{\xi}(p)=\xi \lambda\left[\int_0^{p^2} dq^2 (1-\frac{q^2}{p^2}) S_B(q^2)-
\int d \om \sigma_B(\om)(1+ln\frac{\om}{\mu_t^2})\right] \, .
\ee

Performing the same game  for the part of selfenergy that contributes to the renormalization wave function 
we get
\be
a^{MS}_{\xi}(p)=\xi \lambda\left[\int_0^{p^2} dq^2 (-a^3+2a^2-1) S_A(q^2)-
\int \om \sigma_B(\om)(7/6+ln\frac{\om}{\mu_t^2})\right] \, .
\ee
with $a=q^2/p^2$.
 
The above expressions represent renormalized inverse quark propagators in the MS bar renormalization scheme where
$\mu_t$ stands for the t'Hooft renormalization scale. 
All other renormalization schemes differ by finite constant pieces added or subtracted from equations above.

In MOM scheme we have
\be 
a(p,\mu)=a^{MS}(p)-\Re a^{MS}(\mu)=a(p)-\Re a(\mu) \, , 
\ee
sending $\Re a^{MS}(\mu)$ into the counterterm of $Z$. For the the spacelike value of subtraction point $\mu$,
 the symbol for the real part can be omitted.

\section{The BSE}

 The BSE vertex function for the spin-one bound state composed of a quark-antiquark pair can be expressed as a linear combination of eight transverse trace-orthogonal Lorentz covariants. Of these, only two are employed:

\bea 
\Gamma_1^{\mu}(q,Q)&=&(\gamma^{\mu}-\frac{\not Q Q^{\mu}}{Q^2})\Gamma_1(Q,q)= P_1^{\mu} \Gamma_1(Q,q)
\nn \\
\Gamma_5^{\mu}(q,Q)&=&(q^{\mu}-\frac{q.Q \,Q^{\mu}}{Q^2})\Gamma_5(Q,q)= P_2^{\mu}
\Gamma_2(Q,q)
\eea 
Our limited selection is justified in hindsight and proves sufficient for evaluating meson masses.  

For the $h_b$ mesons, we achieve the same by placing $\gamma_5$ in front of projectors $P_i$, which are used to define
two different Lorentz scalars $\Gamma_1(Q,q)$ and $\Gamma_5(Q,q)$in our approximation. In order 
to avoid notational cluttering we use the same name for the components of vector and  axial vector mesons.

After the projection, we analytically integrate over the 3d space angles in the rest frame 
of the meson, where $Q=(M,0)$. To numerically integrate the BSE, we work in the complex momentum Euclidean space, where the relative momentum of the quark-antiquark pair is identified with integration momentum 
$q=(q_0,\vec{q})\rightarrow (iq_4,\vec{q})$, while  the total momentum remains "unrotated" due to the  requirement: $Q^2=M^2>0$. 
This quite standard procedure  requires knowledge of the quark propagator for  complex arguments.
\be
q_{\pm}^2=-q^2_E\pm i q_4 M+\frac{M^2}{4} 
\ee
for which purpose  representation (\ref{srq}) turns to be particularly useful.

Thus for instance the product $S_A(+)S_A(-)$ is calculated through
\be
S_A(+)S_A(-)=(\Re S_A)^2+(\Im S_A(-))^2
\ee
where 
\bea
\Re S_A&=&\int do \frac{\sigma_A(o)(-p_E^2+M^2/4-o)}{(-p_E^2+M^2/4-o)^2+p_4^2M^2} \, ;
\nn \\
\Im S_A(-)&=&\int do \frac{\sigma_A(o)p_4 M}{(-p_E^2+M^2/4-o)^2+p_4^2M^2}\, ,
\eea
and similarly through the product $S_B(+) S_B(-)$ where  
$\sigma_A$ is replaced by $\sigma_B$ function.

In this way, the  LR BSE is a be three-dimensional integral equation in our approximation.
To reduce the computer time, we  perform a precise fit of the spectral function and evaluate the integrals analytically within the fit. These fits are described at the end of this appendix.    

After 3d  angular integration the coupled system for the BSE 
components reads
\bea \label{ok}
P_1^2\Gamma_1(q_4,q_s)&=&\int_L \Gamma_1(l_4,l_s) K_{1,1}+\int_L \Gamma_5(l_4,l_s) K_{1,5}
\nn \\
P_5^2\Gamma_5(q_4,q_s)&=&\int_L \Gamma_5(l_4,l_s) K_{5,1}+  \int_L \Gamma_5(l_4,l_s) K_{5,5} \, ,
\eea
where the integration symbol stands for
\be \label{ok2}
\int_L=\frac{2\lambda }{\pi}\int_{-\infty}^{\infty} d l_4 \int_0^{\infty} d l_s\int_{0}^{\infty} d\omega\,. 
\ee
The Wick rotation induced mapping $q,Q\rightarrow q_4,q_s$, such that $-q^2=q^2_E=q_4^2+q_s^2; q.Q=i q_4 M; Q^2=M^2$.

It can be shown, that both components of BSE are real for $Y$ quarkonia, while the fifth component can be chosen
purely imaginary for $h_b$ mesons (assuming the first component is taken real).

 We list the  components of the second equation in (\ref{ok}) and describe some details of derivation  bellow. 
The 2 times 2 matrix components that constitute the rhs Eq. (\ref{ok}) for $\Gamma_1$  reads
\bea
K_{1,5}&=&[\pm S_A(+)S_B(-)+ S_A(-)S_B(+)] 
\left\{(l_s^2 [\Omega(0,1)+\xi E(0,1)]\right.
\nn \\
&+&2(l_E^2-l_4q_4)l_s^2{[C(0)-\xi E(0,2)]}
\nn \\
&-&\left.(l_4^2+l_4q_4)[C(1)-\xi E(1,2)]
+C(2)-\xi E(2,2)\right\}
\eea
\bea
K_{1,1}&=&\mp  S_A(+)S_A(-)(2 U+V)+S_B(+)S_B(-) W
 \\
U&=&(C(2)-\xi E(2,2)) 2
\nn \\
&+&(C(1)-\xi E(1,2))(4 l^2+2(l_4+q_4)l_4)
\nn \\
&+&(C(0)-\xi E(0,2))( 2l^4+4l^2 q_4l_4+2 q_4l_4^3+(l_4^2-q_4^4)l^2-(l_4-q_4)^2 \frac{Q^2}{2})
\nn \\
V&=&\Omega(0,1)(3l^2-2l_4^2-Q^2)+\xi E(0,1)(-l^2-2l_4^2-Q^/2)
\nn \\
W&=&5\Omega(0,1)+\xi E(0,1)
+2(l_4-q_4)^2(C(0)-\xi E(0,2))
\eea
and the ones that constitute the rhs. of Eq. (\ref{ok}) for $\Gamma_5$  reads
\be
K_{5,5}=\left[3\rho_T(\omega)\Omega(1,1)+\xi\epsilon(1,1)\right]\left[(-l_E^2-M^2/4)S_A(+)S_A(-)\pm
S_B(+)S_B(-)\right]
\ee
\be
K_{5,1}=\left[3\rho_T(\omega)\Omega(1,1)+\xi\epsilon(1,1)\right]\left[S_A(+)S_A(-)\pm
S_B(+)S_B(-)\right] \, ,
\ee
where the upper sign  stands for  vector, while the lower one stands for an axial vector. 
We have drop out the terms which are odd in the fourth component of momenta $l,q$,in the kernel $K(1,5)$.
They should give no contribution for ordinary normal state quarkonium (Apart of normal time parity states there can be
abnormal parity states \cite{CUT1900}- the excitations in relative time. We do not consider these exotic states in this paper).

The reader should note that the square of projectors differ for different parity mesons, e.g. $\frac{Tr}{4}P_1^2=3$ for vectors , but $\frac{Tr}{4}P_1^2=-3 $ for axial-vectors.
 It  is identical for   both  parity mesons, explicitly we have $\frac{Tr}{4}P_5^2=-q_s^2$ in our coordinate system.
Functions $\Omega,E$ and $C$  are the following shorthand notations  for the angular integral over the 3d subspace of spatial momentum. 
\be \label{ang}
\Omega(N,M)=I(N,M;\omega) \,\, ; \,\, E(N,M)=I(N,M;m_{\epsilon}^2) \, ;
\ee
\be
C(N)=[\int_{-1}^{1} dz \frac{[z l_s q_s]^N}{[(l-q)^2-\om](l-q)^2}]_{reg}
=\frac{1}{\omega}(I(N,1;\omega)-I(N,1;m_{\epsilon}^2) \, ;
\ee
where  $q_s(l_s)$ represents the norm of the 3d space momentum $\vec{q}(\vec{l})$     
and  the infrared  mass regulator $m_{\epsilon}$ is introduced in the function E and $C$.
Although the integrals  (\ref{ang}) are identical in our approximation, we keep the labeling different in order 
to keep the track of their  different  origin as they come from the transverse or longitudinal gauge term.  The integrals for the specific values $N$ and $M$ are listed in the next appendix (\ref{AAI}).

In order to keep our shorthand notation (\ref{ok2}) correct, the reader is kindly asked to add a trivial delta function in front of the function E, such that 
$E(N,M)\rightarrow \delta(\omega-m_{\epsilon}^2) E(N,M)$. 
The regulator mass is used in all angular integrals stemming from the
gauge fixing term.

In what follow we illustrate some details of our derivation for the case of diagonal matrix kernel $K_{5,5}$,
for which the evaluation is particularly  short. Applying $P_5$ projector cancels out the timelike component of $l.q$ product, explicitly it gives
\be
[l.q-\frac{l.Q q.Q}{Q^2}]=-l_s q_s z
\ee
in the numerator of BSE (here $q$ is the external relative momentum, $l$ is the integration one).
Making the trace one gets for  projected rhs. of BSE the following expression
\bea
&&\simeq \int{d^4l} \Gamma_5(l,Q)(-l_s q_s z)\left[(l^2-\frac{Q^2}{4})S_A(+)S_A(-)\pm S_B(+)S_B(-)\right]
\nn \\
&&\left[\frac{3\rho_T(\omega)}{(l-q)^2-\omega}-\frac{\xi}{(l-q)^2-m_{\epsilon}^2}\right]+... \, ,
\eea
where the dots represent contribution stemming from other components of BSE. After performing the Wick rotation  and  performing angular integrations one can identify the diagonal ${5,5} $ element of the integral kernel $K$. The derivation of remaining kernel components follows the same steps, but  related algebra is  more involving in these cases.

\section{ Assorted Angular integrals}
\label{AAI}

Here we sort all necessary  angular integrals. Let us label
the function which implicitly depends on the two variables $a$ and $b$ defined as: 
\be
 a=(l_4-q_4)^2+l_s^2+q_s^2+o\; ;\; b=2 q_s l_s
\ee
in the following way
\be
I(N,M;o)=\int^1_{-1} dz \frac{(q_s l_s z)^N}{[-(l_E-q_E)^2-o]^M}  \,.
\ee

The list of results for relevant integers $N,M$ reads:
\be
I(0,1;o)=\frac{1}{b} ln\left(\frac{o_-}{o_+}\right)
\ee
\be
I(0,2;o)=\frac{2}{o_-o_+}
\ee

\be
I(1,1;o)=1+\frac{a}{2b} ln\left(\frac{o_-}{o_+}\right)
\ee
\be
I(1,2;o)=\frac{1}{2b} ln\left(\frac{o_-}{o_+}\right)+\frac{a}{o_-o_+}
\ee

\be
I(2,1;o)=\frac{a}{2}+\frac{a^2}{4b} ln\left(\frac{o_-}{o_+}\right)
\ee
\be
I(2,2;o)=\frac{1}{2}\left[1+ \frac{a}{2b}ln\left(\frac{o_-}{o_+}\right)+\frac{a^2}{o_-o_+}\right]
\ee
where we have defined the variable  $o_{\pm}=a\pm b$..

All problematic terms in BSE kernel involve the integral of the form
$I(N,2;o=0)$ which would  lead to divergent integrals  due to the double  integration over the variable  $l_s$ and $l_4$. 

Integrals with $M=2$ can be easily derived from those with $M=1$ by differentiation with respect to the variable $o$.

\subsection{Fit of DSE solution for the quark propagator}

The Cauchy distribution 
$C(z;a,b)=\frac{1}{(z-a)^2+b)}$ and its product with monomial i.e. $z C(z;a,b)$ are used to make a fit of spectral functions. Such functions are particularly useful since the  only a few 
Cauchy distributions are needed to achieve precise fits.
Furthermore, within the fit, the analytical continuation of the propagator is known for 
complex arguments.

Basic dispersion relation  over the Cauchy distribution reads
\bea
&&I_B(s;a,b)=\int_0^{\infty}dz \frac{C(z;a,b)}{s-z}=
\nn \\
&&=C(s;a,b)\left[\frac{1}{\sqrt{b}}(\pi-tg^{-1}(\frac{\sqrt{b}}{a})(s-a)+\frac{1}{2} ln(\frac{s^2}{a^2+b})\right]
\nn \\
&&\int_0^{\infty}dz \frac{z C(z;a,b)}{s-z}=s I_B(s;a,b)-K
\eea
where $s$ stands for complex square of momentum with non  zero imaginary part. 
The real constant $K$ reads
\be
K=\frac{1}{\sqrt{b}}(\pi/2+tg^{-1}(\frac{a}{\sqrt{b}})\,.
\ee

The following fit for the spectral function was used
 \be
 \sigma_A(o)=c_1C(o;a,b)
 \ee
 with the coefficients $ c_1=0.041 b$  and dimesionfull parameters $a=24 GeV^2$ and  $b=50 GeV^4$. 

For the second spectral function 
we achieve very goof fit with the following combination
\bea
\sigma_B(o)&=&c_2C(o;a,b)+c_3 o C(o;a,b)
 \\
 c_2&=&(1-x) 0.21 b   \, ;\,  c_3=x 0.21 b/a
 \eea
with the same parameters $a,b$ as above and $x=0.45$. These fits are compared with the numerical solution 
of DSE in the Fig. \ref{zuva}.

As an added benefit, we can accurately estimate the precision of our numerical P integration, which is used to calculate spectral deviation. The numerical values of the integrals differ from the exact values by the fifth digit.
This indicates that the spectral deviation has a systematic origin in the approximation/truncation of DSEs and is not caused by numerical noise.

\subsection{Auxiliary normalization, the eigenvalue function}

The eingevalue function $N$ that  we have used in our numerics  involves a certain, albeit unphysical norm of the first component BSE function. For purpose of completeness we show it here:
\be \label{normalization}
N=\int_{-\infty}^{\infty}d q_4 d q_s \Gamma_1(q_4,q_s) \frac{\sqrt{q_4}\sqrt{q_s}}{(q_4^2+1)(q_s^2+1)(1+q_4 M/2+q_s^2)^{1/2}} \, 
\ee
with all momenta in units of GeV.

The CBSE components have been normalized such that  $N$ lies within the interval 
$(0,99,1.02)$. The closer $N$ to one is, the smaller the difference between two consecutive iterations.
The sample of searches is shown in the figure \ref{last}.

 {\mbox{}}
\\
\begin{figure}[ht]
\centerline{\includegraphics[width=8.0cm]{eigen.eps}}
\caption{\label{last}Sample of numerical search. The eigenvalue auxiliary function $N$ for CBSE2 and the integral error of iterations $\sigma$ for all approximations are   shown against the total momentum of the quark-antiquark pair.}
\end{figure}
\\
 {\mbox{}}


%
\end{document}